\documentstyle[11pt]{article}

\newcommand{\be}{\begin{equation}}
\newcommand{\ee}{\end{equation}}
\newcommand{\bea}{\begin{array}}
\newcommand{\ea}{\end{array}}
\newcommand{\beqa}{\begin{eqnarray}}
\newcommand{\eeqa}{\end{eqnarray}}
\newcommand{\bean}{\begin{eqnarray*}}
\newcommand{\eean}{\end{eqnarray*}}
\newcommand{\eqn}[1]{(\ref{#1})}

\def\up#1{\leavevmode \raise.16ex\hbox{#1}}

\def\sqr#1#2{{\vcenter{\vbox{\hrule height.#2pt
        \hbox{\vrule width.#2pt height#1pt \kern#1pt
          \vrule width.#2pt}
        \hrule height.#2pt}}}}

\newcommand{\gapproxeq}{\lower .7ex\hbox{$\;\stackrel{\textstyle >}{\sim}\;$}}
\newcommand{\lapproxeq}{\lower .7ex\hbox{$\;\stackrel{\textstyle <}{\sim}\;$}}

\newcounter{appendice}

\def\pr{{\it Phys. Rev.}\ }

\def\cqg{{\it Class. Quantum Grav.}\ }

\def\grg{{\it Gen. Relativ. Grav.}\ }

\def\apj{{\it Ap. J.}\ }
\def\aa{{\it Astron. Astrophys.}\ }

\def\prep{{\it Phys. Rep.}\ }

\def\thebibliography#1{{\bf REFERENCES\markboth
 {REFERENCES}{REFERENCES}}\list
 {[\arabic{enumi}]}{\settowidth\labelwidth{[#1]}\leftmargin\labelwidth
 \advance\leftmargin\labelsep
 \usecounter{enumi}}
 \def\newblock{\hskip .11em plus .33em minus -.07em}
 \sloppy
 \sfcode`\.=1000\relax}

\begin{document}

\title{\hfill $\mbox{\small{
$\stackrel{\rm\textstyle  hep-th//yymmdd}
{\rm\textstyle DSF-55/97\quad}
{\rm\textstyle DSF-US 11/97\quad} $}}$
\\[1truecm] Cosmological Waveguides for Gravitational Waves}
\author{G. Bimonte,$^{1}$ S. Capozziello,$^{2}$ V. Man'ko, $^{3,4}$ and
G. Marmo $^{1}$}
\maketitle
\thispagestyle{empty}

\begin{center}
{\it 1)}  Dipartimento di Scienze Fisiche, Universit\`a di Napoli,\\ Mostra
d'Oltremare, Pad.19, I-80125, Napoli, Italy; \\ INFN, Sezione di Napoli,
Napoli, ITALY.\\ {\it 2)} Dipartimento di Scienze Fisiche "E.R.
Caianiello", Universit\'a degli Studi di Salerno,\\ I-84081 Baronissi
(Salerno), ITALY.\\ {\it 3)} Osservatorio Astronomico di Capodimonte,\\ V.
Moiariello 16, I-80131, Napoli, ITALY.\\ {\it 4)} P. N. Lebedev Physical
Institute, Leninsky Pr. 53, Moscow 117924, RUSSIA.\\

\small e-mail: {\tt bimonte, gimarmo, capozziello@napoli.infn.it }\\
\end{center}

\begin{abstract}

We study the linearized equations describing the propagation of
gravitational waves through dust. In the leading order of the WKB
approximation, dust behaves as a non-dispersive, non-dissipative medium.
Taking advantage of these features, we explore the possibility that a
gravitational wave from a distant source gets trapped by the  gravitational
field of a long filament of galaxies of the kind seen in the large scale
structure of the Universe. Such a waveguiding effect may lead to a huge
magnification of the radiation flux from distant sources, thus lowering the
sensitivity threshold required for a successful detection of gravitational
waves by detectors like VIRGO, LIGO and LISA.

\end{abstract}

\bigskip

\bigskip

\vspace{2. cm}

PACS: 04.25, 04.30-w, 98.62.Sb

\newpage
\section{Introduction}
Detecting gravitational waves is one of the greatest challenges for modern
physics both from the theoretical and the experimental points of view
\cite{abramovici,bradaschia,bender}. The interest of this enterprise arises
from the fact that, in case of success, it will open an entirely new window
on the Universe and, as our past experience with radio astronomy teaches,
this might have a profound impact on our vision of the Universe.
Unfortunatly, the detection of a gravitational wave has turned out to be a
very difficult task and as of now, almost fourty years after the early
pioneering work of Weber \cite{web}, no one has succeded yet. The main
source of the difficulties is the fact that most of the known astrophysical
sources should radiate exceedingly small amounts of gravitational
radiation. The strongest sources should be massive, highly non-sperical
systems with large internal kinetic energies \cite{thorne}, like colliding
and coalescing black holes and neutron stars.  Sources of this type last
for a short time and should also be very rare, with the consequence that it
will be necessary to search large regions of the Universe in order to have
a reasonable event rate. For sources in the high frequency band ($1\div
10^4$ Hz), accessible to ground based interferometers like VIRGO and LIGO,
a rate of several coalescence events per year requires extending the search
at distances from the Earth of the order of $10^2\div 10^3$ Mpc. At such
large distances even the dimensionless amplitudes $h$ of the waves from
such strong sources should be in the range of $h\sim 10^{-21}
\div 10^{-22}$, which still represents an extremely small signal. There are
hopes that the LIGO/VIRGO collaborations
\cite{abramovici},\cite{bradaschia} will be able to reach or at least
closely approach the sensitivity needed for the detection of these signals.
The same considerations apply to the sources in the low-frequency band
($10^{-4}\div 1$ Hz) accessible to the forthcoming orbiting LISA detector.
Examples of such sources from distant galaxies, at distances of the order
of one Gpc from the Earth, are coalescing massive black holes, or compact
bodies spiraling into one of them. For a more thorough discussion of these
issues and an updated guide to the bibliography we refer the reader to the
recent paper by Thorne \cite{thor2}.

This being the state of things, it is our opinion that any mechanism
leading to an amplification of the signal from a cosmological source of
gravitational waves, thus making its detection easier, should be very
welcome. An example of such a mechanism, well established by now in the
realm of optical and radio-wave astronomy, is provided by gravitational
lensing. In fact, the possibility that the gravitational radiation from
distant binary neutron stars may get magnified due to microlensing by
compact bodies has been recently considered in the literature \cite{wang}.
In this paper we propose a different lensing effect: we shall argue that
the large scale galactic gravitational fields generated by the filamentary
distributions of galaxies seen in the large scale structure of the
Universe, primordial cosmic strings evolved in today observable structures,
may act as waveguides for gravitational waves. Since a gravitational wave
interacts very weakly with practically all kinds of matter, a wave trapped
in the interior of such a filament might be able to traverse cosmological
distances without being absorbed or scattered in a significant way, and
most importantly without undergoing the $1/R$ attenuation that charaterizes
propagation in vacuum. If a significant fraction of the radiation from
distant sources is waveguided to us in this way, this might lower in an
appreciable way the minimum sensitivity needed for detecting the signal.

The idea of gravitational conductors, waveguides and circuits is not new
per se. For example, it was considered in \cite{press} where it was based
on the hypothetical existence of a material having a large shear (the
"respondium") and capable of reflecting a gravitational wave. Our approach
is completely different and not so exotic. Starting from standard General
Relativity, we show that a filamentary distribution of dust is capable of
trapping in its interior a gravitational wave: this result will be derived
below by applying the (lowest-order) WKB approximation to the linearized
Einstein's equations describing the propagation of a weak gravitational
wave in a region of space filled with dust. Within these approximations,
the trapping of a gravitational wave appears completely analogous to that
of a ray of light inside a guide with a specific refraction index profile.
Being based on the ray approximation, our approach cannot take into account
interference effects, which should be negligible in the high frequency band
accessible to VIRGO/LIGO as well as in the low frequency band accessible to
LISA. In case they could not be neglected, one could resort to an
approximation of the type of the Fock-Leontovich paraxial approximation
\cite{fock}, which, while using the short-wave approximation, allows also
to describe interference.

A waveguiding effect by cosmological structures analogous to that described
in this paper was shown to be possible in Ref. \cite{cap} for the case of
electromagnetic waves and in Refs.\cite{dodma,dodma1} for massless
particles, using the Fock-Leontovich approximation. In Ref.\cite{cap}, in
particular, this mechanism was used to explain in a simple way the huge
luminosity of quasars, as compared to their high redshift. They would not
be exotic objects hiding in their interior some unknown engine that
releases enormous amounts of energy; rather, they might just be
protogalaxies whose luminosity is preserved by a waveguide. Furthermore,
certain gravitational lensing effects such as``twin" and ``brother" quasars
\cite{dodonov} could be explained in a similar manner.

The paper is organized as follows. In Sect. 2 we present the linearized
theory of gravitational radiation in a region of space-time filled with
dust. We show that, in the leading WKB approximation, dust behaves as a
non-dispersive and non-dissipative medium. In Sect. 3, we use this
approximation to study the possibility that a gravitational wave, coming
from a distant source, could be trapped by the gravitational field of a
long filamentary structure, thus yielding to a waveguide effect. We
estimate the focalization length, the amount of luminosity preserved, and
the effective feasibility of the effect. A discussion of the results and
the conclusions are drawn in Sect. 4.

\section{Linearized Theory}

In this Section, we shall derive the linearized equations describing the
propagation of a weak gravitational wave through a region of space-time
filled with dust. \footnote{We shall follow in this paper the notations and
conventions of \cite{wald}. In particular, the signature of the metric is
chosen to be $(-+++)$. Moreover $A_{(bc)}:=\frac{1}{2}(A_{bc}+A_{cb})$.}.
The same problem was discussed in Ref.\cite{ehl}, but the equations derived
below are simpler than those of \cite{ehl}. Before the passage of the wave
we thus have an unperturbed background $(g_{ab}, u_a,
\rho)$, where $g_{ab}$ is the space-time metric while $u_a$ and $\rho$
denote, respectively, the field of velocities and the density of the dust.
The background is assumed to satisfy the exact Einstein's equations in the
presence of dust:
\be
G_{ab}= \kappa \rho u_a u_b~,\label{eins}
\ee
where $\kappa= 8 \pi G/c^4$, $G$ being Newton's constant, while $G_{ab}$ is
the Einstein's tensor
\be
G_{ab}=R_{ab}- \frac{1}{2} g_{ab} R ~.
\ee
We also have
\be
\rho > 0~,
\ee
while
\be
u_a u^a = -1~.\label{usq}
\ee
We recall that the conservation law for the stress-energy of the dust,
combined with the normalization condition (\ref{usq}) implies that the
velocities $u^a$ ~satisfy the geodesic equation:
\be
u^b \nabla_b u^a=0~,
\ee
while the density $\rho$ satisfies the conservation equation:
\be
\nabla_a(\rho u^a)=0~.
\ee
We now imagine perturbing the background solution by a small perturbation
$(g_{ab}+\hat{g}_{ab},u_a+\hat{u}_a,\rho+ \hat{\rho})$. Plugging the
perturbed solution into Einstein's equations and retaining all the terms up
to first order in the perturbation, we get the following linearized
equations:
\be
\hat{G}_{ab}=\kappa[\hat{\rho}u_a u_b + \rho(\hat{u}_a u_b + u_a \hat{u}_b)]~,\label{linein}
\ee
where
\beqa
\hat{G}_{ab} = &-&\frac{1}{2} \nabla^c \nabla_c \hat{\gamma}_{ab} +
\nabla_{(a}\nabla^c \hat{\gamma}_{b)c}-\frac{1}{2}g_{ab}\nabla^c \nabla^d \hat{\gamma}_{cd}\;
\nonumber \\
& + &R^{c~~~d}_{~ab}
\hat{\gamma}_{cd} + R_{(a}^{~~c} \hat{\gamma}_{b)c}-\frac{1}{2}
R\hat{\gamma}_{ab} + \frac{1}{2}g_{ab}R^{cd}\hat{\gamma}_{cd}~.\label{hatg}
\eeqa
In this equation $\nabla_a$ and $R_{abc}^{~~~~d}$ represent, respectively,
the covariant derivative and the Riemann tensor relative to the background
metric $g_{ab}$. Indices are raised and lowered using $g_{ab}$ and its
inverse, while $\hat{\gamma}_{ab}$ is defined as
\be
\hat{g}_{ab}=\hat{\gamma}_{ab}-\frac{1}{2}g_{ab}\; \hat{\gamma}~,
\ee
where
\be
\hat{\gamma} = g^{ab} \hat{\gamma}_{ab}~.
\ee
 We have to add to Eq. (\ref{linein}) the equation obtained linearizing
the analogue of the normalization condition Eq. \eqn{usq} for the perturbed
velocity field $u_a +\hat{u}_a$:
\be
2 u^a \hat{u}_a = \hat{g}_{ab} u^a u^b~.\label{linnor}
\ee
In deriving this equation, we used:
\be
\hat{g}^{ab}=-g^{ac}g^{bd}\hat{g}_{cd}~
\ee
and
\be
\hat{u}^a =g^{ab} \hat{u}_b-g^{ab} u^c \hat{g}_{bc}~.
\ee
Thus, Eqs. \eqn{linein} and \eqn{linnor} are the equations that describe
the propagation of a weak gravitational wave in a medium made of dust. We
refer the  reader to Ref. \cite{ehl} for a proof of their linear stability.
Let now $h^a_b$ be the projector onto the hyperplane orthogonal to the
4-velocity $u^a$:
\be
h^a_b= \delta^a_b + u^a u_b~. \label{proj}
\ee
Upon multiplying Eq. \eqn{linein} by $h^a_e h^b_f$, we get:
\be
h^a_e h^b_f \; \hat{G}_{ab}=0~.\label{evo}
\ee
Upon multiplying now Eq. \eqn{linein} by $u^a u^b$ and using Eq.
\eqn{linnor} to eliminate $u^a \hat{u}_a$, we also get:
\be
\kappa \hat{\rho} = \hat{G}_{ab}u^a u^b + \kappa \rho \hat{g}_{ab}u^a u^b~.\label{hatr}
\ee
Finally, by multiplying Eq. \eqn{linein} by $u^b$ and using again Eq.
\eqn{linnor} together with  Eq. \eqn{hatr}, we get:
\be
\kappa \rho \hat{u}_a= -(\hat{G}_{cd}u^c u^d \; u_a + \hat{G}_{ab}u^b)
-\frac{\kappa}{2} \rho \; \hat{g}_{cd} u^c u^d\; u_a ~.\label{hatu}
\ee
Equations \eqn{evo}, \eqn{hatr} and (\ref{hatu}) are completely equivalent
to Eqs. (\ref{linein}) and $\eqn{linnor}$. This can be seen by plugging Eq.
\eqn{proj} in Eq. \eqn{evo} and then taking the terms involving the velocities
$u_a$ on the right hand side: upon using Eqs. \eqn{hatr} and \eqn{hatu},
one then recovers the right hand side of Eq. \eqn{linein}. Finally, Eq.
(\ref{linnor}) can be obtained back from Eq. \eqn{hatu} by simply
multiplying it by $u^a$.

Now, Eq. \eqn{evo} is the only evolution equation governing the
perturbation, while Eqs. \eqn{hatr} and \eqn{hatu} just represent
constraints. It is quite remarkable that the evolution equations depend
only on the perturbation of the metric, while the constraints are
explicitly solved with respect to the perturbations of the velocities and
density. Thus, we focus on Eqs. \eqn{evo}: they can be simplified a bit by
a suitable choice of gauge. Under an infinitesimal diffeomorphism
\be
x^{\prime a}=x^{a}-\xi^{a}~,
\ee
$\hat{\gamma}_{ab}$ transforms as:
\be
\hat{\gamma}_{ab}^{\prime}=\hat{\gamma}_{ab} + 2 \nabla_{(a} \xi_{b)}- g_{ab} \nabla^c \xi_c~.
\label{diff}
\ee
Using the method discussed in \cite{ehl}, it is easy to prove that by means
of a suitable choice of $\xi_a$ it is always possible to achieve the
so-called transverse gauge:
\be
\nabla^b \hat{\gamma}_{ab}=0~.\label{gauge}
\ee
We notice that this transversality condition does not fix completely the
gauge, as it is preserved by any infinitesimal diffeomorphism such that:
\be
\nabla^a \nabla_a \xi_ b + R_b^{~a}\xi_a=0~.\label{resi}
\ee
In the transverse gauge, the second and third terms in the expression of
$\hat{G}_{ab}$ in Eq. \eqn{hatg} drop out and thus we are left with the
following equations:
\be
h^a_e h^b_f \;\left(\frac{1}{2} \nabla^c \nabla_c \hat{\gamma}_{ab}
-R^{c~~~d}_{~ab}
\hat{\gamma}_{cd} - R_{(a}^{~~c} \hat{\gamma}_{b)c}
+ \frac{1}{2} R\hat{\gamma}_{ab} -
\frac{1}{2}g_{ab}R^{cd}\hat{\gamma}_{cd} \right)=0~,\label{evo2}
\ee
that are to be solved together with Eq. \eqn{gauge}.

A further simplification arises if the wavelength $\lambda$ of the
perturbation is much smaller than all the typical lengths that characterize
the background. We thus assume that the distances over which the velocities
$u_{a}$ vary in an appreciable way and the average radius of curvature of
the background $|R_{abcd}|^{-1/2}$ are both much larger than $\lambda$.
Under these conditions, it is legitimate to make the WKB ansatz:
\be
\hat{\gamma}_{ab}(x,\epsilon)= {\rm Re} \{f_{ab}(x,\epsilon)\;\exp(i/\epsilon\;S(x)) \}~,
\label{wkb}
\ee
where $f_{ab}(x,\epsilon)$ is a slowly varying amplitude having the
asymptotic expansion (for $\epsilon \rightarrow 0$):
\be
f_{ab}(x,\epsilon)= \sum_{n=0}^{\infty} \left(\frac{\epsilon}{i}\right)^n
f^{(n)}_{ab}(x)~.
\ee
Plugging this expansion in Eq. \eqn{evo2} and equating to zero the
coefficients of each power of $\epsilon$, one gets a system of recursive
equations for the phase $S(x)$ and for the amplitudes $f^{(n)}_{ab}(x)$. To
order $\epsilon^{-2}$ we get:
\be
(g^{ab} l_a l_b) \; (h^c_e h^d_f f^{(0)}_{cd})=0~, \label{lowest}
\ee
where $l_a$ is the wave vector:
\be
l_a := \partial_a S~.\label{ell}
\ee
As for the transversality condition Eq. \eqn{gauge}, it gives to order
$\epsilon^{-1}$:
\be
l^b  f^{(0)}_{ab}=0~.\label{gzero}
\ee
Equation \eqn{lowest} admits two types of solutions, depending on which of
the two factors between the brackets vanishes. Consider first:
\be
h^c_a h^d_b f^{(0)}_{cd}=0~.\label{one}
\ee
This equation implies that $f^{(0)}_{ab}$ is of the form:
\be
f^{(0)}_{ab}= v_{(a} u_{b)}:=\frac{1}{2}(v_a u_b +u_a v_b)~,\label{one2}
\ee
where $v_a$ in an arbitrary vector field. Upon multiplying both sides of
the above equation by $2 l^b$, and imposing Eq. \eqn{gzero}, one finds that
$v_a$ must be proportional to $u_a$, $v_a = v u_a$. But then, multiplying
Eq. \eqn{one2} by $l^a l^b$, one gets that $f^{(0)}_{ab}$ can be different
from zero only if $\omega:=-l^a u_a =0$. This means that the wavefronts are
time-like surfaces, and then this type of solutions cannot describe a
gravitational wave. For this reason, we shall not consider anymore this
case in the rest of the paper.

The second class of solutions of Eq. \eqn{lowest} is characterized by
phase-functions $S(x)$ satisfying the eikonal equation:
\be
g^{ab} (\partial_a S)\;( \partial_b S)=0~.\label{eik}
\ee
The wavefronts are thus null hypersurfaces (we exclude the possibility of
singular wavefronts and thus assume that $l_a$ does not vanish at any
point) describing waves that propagate through the dust with the speed of
light. We recall that the integral curves $x^a(v)$, called rays, of the
null vector field $l_a$:
\be
\frac{d x^a}{dv}=(g^{ab} \partial_b S) (x(v)),\label{rays}
\ee
with $S(x)$ a solution of the eikonal equation, are null geodesics of the
background metric $g_{ab}$ providing bicharacteristics of Eq. \eqn{evo2}.

By looking at the next order in the formal expansion of Eq. \eqn{evo2}, at
order $\epsilon^{-1}$ one gets a continuity equation for the amplitude
$f^{(0)}_{ab}(x)$:
\be
h^c_a h^d_b ( \nabla_l +  \theta )f^{(0)}_{cd}=0~, \label{zero}
\ee
where we have set
\be
\nabla_l:= l^a \nabla_a~,~~~~~\theta:= \frac{1}{2}\nabla_a  l^ a~.
\ee
Now, Eq. \eqn{zero} is equivalent to
\be
(\nabla_l +  \theta )f^{(0)}_{ab}= v_{(a} u_{b)}~,\label{ezero}
\ee
where $v_a(x)$ is again an arbitrary vector field. Eq. \eqn{gzero} together
with the geodesic condition satisfied by $l_a$, $\nabla_l l_a=0$, imply
that $v_a$ must be identically zero. To see this, multiply both sides of
Eq. \eqn{ezero} by $ 2 l^a$. The left-hand side is then seen to vanish:
$$
2 l^b(\nabla_l +  \theta )f^{(0)}_{ab}= 2 \nabla_l (l^b f^{(0)}_{ab})=0~,
$$
and thus we have for the right-hand side:
\be
0=2 l^b v_{(a} u_{b)} =(l^b v_b) u_a - \omega \; v_ a~.\label{lambda}
\ee
This equation implies that $v_a$ must be proportional to $u_a$: $v_ a
=v \; u_a$; but then it easily follows that it must vanish because
after multiplying Eq. \eqn{ezero} by $l^a l^b$ we get
$$
0= v\; \omega^2~,
$$
which implies that $v$ vanishes, because $\omega$, being the time-component
of a non-vanishing null vector, is different from zero. We conclude that
the amplitude $f^{(0)}_{ab}$ satisfies the transport equation:
\be
(\nabla_l +  \theta)f^{(0)}_{ab}=0~.\label{trans}
\ee
It follows from this equation that the expansion of $G_{ab}$ vanishes up to
order $\epsilon^{-1}$ included and thus, according to Eqs. \eqn{hatr} and
\eqn{hatu}, the dust suffers no density or velocity perturbations
to this order.

As it is usual in the WKB approximation, Eq. \eqn{trans} converts into an
{\it ordinary} first-order differential equation for the restriction
$f^{(0)}_{ab}(x(v))$ of the amplitude $f^{(0)}_{ab}(x)$ to the rays
\eqn{rays} and thus its solutions are uniquely determined by the knowledge
of $f^{(0)}_{ab}$ on a surface $\Sigma$ cutting each ray at one point. We
can further simplify Eq. \eqn{trans} by a suitable choice of a tetrad
field, which we now describe. We take $u^a$ and $k^a:=h^a_b l^b$ as the
first two elements of the tetrad, while the remaining two, that we call
$e^{1}_a,~e^{2}_a$ are defined as follows. We pick up an arbitrary point on
each ray and choose for $e^{1}_a,~e^{2}_a$ there two arbitrary space-like
unit vectors such that:
\be
e^{i}_a u^a = e^{i}_a l^a =0~,~~~e^{1a}e^{2}_a=0~.
\ee
The vectors $e^{i}_a$ are then transported along the ray according to the
``quasiparallel" transport rule \cite{breu} given by the equation
\be
\nabla_l e^{i a}=-e^{i b}\nabla_l q^a_b~,\label{rule}
\ee
where $q^a_b$ is the orthogonal projector onto the plane spanned by $u^a$
and $k^a$:
\be
q^a_b=-u^a u_b + \frac{1}{\omega^2}k^a k_b~.
\ee
In \cite{breu} it is shown that this rule of transport preserves inner
products and that $e^{i}_a$ remain orthogonal to $u^a$ and $k^a$ all along
the ray. Using the tetrad so defined, one can see that the most general
amplitude $f^{(0)}_{ab}$ satisfying the gauge condition Eq. \eqn{gzero} can
be decomposed as
\be
f^{(0)}_{ab}= A_+ e^+_{ab} +  A_{\times} e^{\times}_{ab}+ A_{tr}
e^{tr}_{ab} + B_i e^i_{(a} l_{b)}+ B l_a l_b~,\label{deco}
\ee
where
$$
e^+_{ab}=e^1_a e^1_b-e^2_a e^2_b~,~~~~~~~~e^{\times}_{ab}=e^1_a e^2_b +
e^2_a e^1_b~,
$$
\be
e^{tr}_{ab}=e^1_a e^1_b + e^2_a e^2_b~.\label{emod}
\ee
$e^+_{ab}$ and $e^{\times}_{ab}$ represent the usual traceless transverse
(T-T) modes of a gravitational wave, while $e^{tr}_{ab}$ is a transverse
mode with non-vanishing trace; as for the last two terms in Eq. \eqn{deco},
they represent longitudinal modes. The coefficients
$A_+,~A_{\times},~A_{tr},~B_i$ and $B$ satisfy a simple transport equation.
Upon inserting Eq. \eqn{deco} in Eq. \eqn{trans} and taking its trace, we
find:
\be
(\nabla_l + \theta)A_{tr}=0~.
\ee
Contracting now Eq. \eqn{trans} with any of the traceless modes appearing
in Eq. \eqn{deco}, in view of the fact that they are orthogonal to each
other, and observing that $e^i_a
\nabla_l e^{ja}=0$, which is a consequence of the transport rule
Eq. \eqn{rule}, one obtains:
\beqa
(\nabla_l +  \theta)A_+~&=&\;(\nabla_l +
\theta)A_{\times}\;=\;\nonumber\\
&=&(\nabla_l +  \theta)B_i\;=\;(\nabla_l +
\theta)B\;=0~.\label{amp}
\eeqa
We now show that by means of a gauge transformation of the type \eqn{resi}
it is always possible to achieve that $A_{tr}$, $B_i$, and $B$ vanish on
some initial surface $\Sigma$ and, due to the above transport equations,
this implies that they vanish all along the rays crossing $\Sigma$. For
this purpose, we consider a family of vector fields $\xi(x,\epsilon)$ of
the form
\be
\xi_a(x,\epsilon)= e^{i/\epsilon\; S(x)} \; \sum_{n=1}^{\infty}
\left(\frac{\epsilon}{i}\right)^n
\xi^{(n)}_a(x)~.
\ee
This ensures that gauge transforming a $\hat{\gamma}_{ab}$ of the form
\eqn{wkb} according to Eq. \eqn{diff} the new $\hat{\gamma}_{ab}$ is still of the form
\eqn{wkb}. In particular, we find for $f^{(0)}_{ab}$
\be
f^{(0)\prime}_{ab}=f^{(0)}_{ab}+ 2 l_{(a}\xi^{(1)}_{b)}-(l^c
\xi^{(1)}_c)g_{ab}~.
\ee
Upon imposing Eq. \eqn{resi} on the field $\xi(x,\epsilon)$, one can see
that the coefficient $\xi^{(1)}_a(x)$ must satisfy the transport equation
\be
(\nabla_l +  \theta)\xi^{(1)}_a=0~.
\ee
The important thing is that the value of $\xi^{(1)}_a$ on $\Sigma$ is
completely arbitrary and it is easy to verify that it always possible to
choose it in such a way that
\be
A^{\prime}_{tr}=B^{\prime}_i=B^{\prime}=0~~~~{\rm on~} \Sigma~.
\ee
Consider now the effective energy-momentum tensor $\hat{T}^{ab}$
\cite{isaa} and the effective ``graviton number" $N^a$, respectively,
defined as
\beqa
\hat{T}^{ab}&=&\frac{1}{4 \kappa}(|A_+|^2+|A_{\times}|^2)\;l^a l^b~,\nonumber\\
N^a&=&\frac{1}{4 \kappa \hbar}(|A_+|^2+|A_{\times}|^2)\;l^a~.
\eeqa
The transport equations for $A_+$ and $A_{\times}$, together with the
geodesic equation for $l_a$, imply that their covariant divergences vanish:
\be
\nabla_a \hat{T}^{ab}=0~,~~~~~\nabla_a N^a=0~.
\ee
Using the identity $\Gamma^b_{ab}=\partial_a g/(2 g)$, where $g=$det
$\parallel g_{ab}\parallel$, we can rewrite the equation $\nabla_a N^a=0$
in the form of a conservation law:
\be
4 \kappa \hbar \nabla_a N^a = \partial_a(|A|^2\;l^a) + |A|^2 l^a
\frac{1}{2g}
\partial_a g=
\frac{1}{\sqrt{-g}}\partial_a(\sqrt{-g} |A|^2 l^a)=0~,\label{cons}
\ee
where $A^2=(|A_+|^2+|A_{\times}|^2)$. Upon integrating this equation on the
4-volume spanned by a ray bundle, and using Gauss' theorem, one proves that
the number of gravitons in the bundle is conserved and
observer-independent.

The conclusion of this analysis is that, in the lowest WKB approximation,
the physical degrees of freedom of a gravitational wave propagating through
dust are represented, as in vacuum, by two traceless-transverse modes,
$e^+_{ab}$ and $e^{\times}_{ab}$. The transport equations for $A_+$ and
$A_{\times}$, Eqs. \eqn{amp}, show that these modes propagate along null
geodesics of the background gravitational field, again as it happens in
vacuum. The dust behaves as a non-dispersive, non-dissipative medium that
preserves the polarization of the wave and the total graviton number in a
ray bundle.

\section{The Waveguiding Effect}

It is well known that, when the WKB approximation is valid, a weak
gravitational wave propagates in vacuum much like an electromagnetic wave.
One is thus led to the expectation that a gravitational wave passing by
some massive body may undergo a lensing effect, much in the same way as it
happens to light. The results of the previous Section show that under many
respects a gravitational wave propagating through dust does not behave
differently than in vacuum. In particular, we saw that dust behaves as a
non-dissipative medium which implies that a gravitational wave can traverse
large amounts of dust without suffering a significant absorption by the
medium. This suggested to us the idea that  the long, filamentary sequences
of galaxies seen in the large scale structure of the Universe may act as
``optical" guides for the gravitational radiation that propagates in their
interior: the gravitational wave from a source placed at one end of such a
filament (or inside it) may get {\it trapped} by the large scale galactic
gravitational field and consequently be able to traverse huge distances
without undergoing the $1/R$ attenuation that characterizes propagation in
vacuum, resulting in a possibly strong magnification of the source for an
observer placed at the other end of the guide. The possibility of an
analogous phenomenon for the electromagnetic radiation was studied in
ref.\cite{cap}.

In this Section we shall study this phenomenon in an highly idealized
situation. We shall consider a point source $S$ of monochromatic
gravitational waves, placed at a large distance from an observer $O$.
Between them, there is a long cylinder ${\cal C}$ filled with dust. We
assume that $S$, $O$ and the cylinder are at rest and that $S$ and $O$ are
roughly aligned with the axis of the cylinder. For simplicity, we imagine
that the distribution of dust inside the cylinder is stationary, while the
density of matter outside the cylinder is negligible. Moreover, we assume
that the background gravitational field $g_{ab}$, generated by the dust, is
weak (but much stronger than the perturbation) and that the relative motion
of the dust is much slower than the speed of light, which is generally the
case in astrophysical situations. Under these conditions the background can
be described using the Newtonian approximation and so there will be a
quasi-Minkowskian coordinate system $\{x^{\mu}\}\equiv
\{ct,x,y,z\}$ where the dust is practically at rest
and the gravitational field $g_{\mu\nu}$ of the background can be expressed
in terms of the Newtonian potential $\Phi$ generated by the dust:
$$
g_{00}=-\left(1+ \frac{2 \Phi}{c^2}\right)~~,~~~~~~~~~
g_{ij}=\left(1-\frac{2
\Phi}{c^2}\right)\delta{ij}~,
$$
\be
~~~~~~~i,j=1,2,3 \label{metric}
\ee
with
\be
|\hat{g}_{\mu\nu}| \ll \frac {2 |\Phi|}{c^2} \ll 1~.\label{weak}
\ee
We assume that the time variation of the potential $\Phi$, due to the
motion of the dust, is so slow that it can be neglected and thus we take
$\Phi$ to be time-independent. We choose the origin and orientation of the
Minkowskian coordinate system such that the axis of the cylinder $\cal C$
coincides with the $z$-axis. Its top and bottom faces have radius $D$ and
have $z$ coordinates respectively equal to $z=0$ and $z=L$. The source $S$
is placed outside the cylinder at the point $P_S$ of coordinates
$(x=a,\;y=0,\;z=-l)$, with $l>0$ and $|a| \ll D$.

According to the WKB approximation developed in the previous Section, the
propagation of the wave emitted by $S$ is described by the eikonal
equation, Eq. \eqn{eik}. Since the background is time-independent, we can
make for the eikonal $S(x)$ the ansatz:
\be
S(t,x,y,z)= k\;\{\bar{S}(x,y,z)-c t\}~,
\ee
where $k=2\pi\nu_0/c $, $\nu_0$ being the frequency of the wave. Plugging
this expression in the eikonal equation \eqn{eik} and using for the
background the expression in Eq. \eqn{metric} we find
\be
\frac{1}{2}\{\;(\partial_x \bar{S})^2 +(\partial_y \bar{S})^2+(\partial_z \bar{S})^2\;\}
+\; \frac{
2 \Phi}{c^2}(x,y,z)=\frac{1}{2}~.\label{eik2}
\ee
While deriving Eq. \eqn{eik2}, we have divided the eikonal equation by $2
g^{ii}$, expanding inverses and products of $g_{\mu\nu}$ to first order in
$\Phi/c^2$. Equation \eqn{eik2} has the same form of the eikonal equation
describing a light-wave propagating in a medium with a refractive index
$n=1-2\Phi/c^2$.

In order to study the propagation of the wave inside the cylinder, we shall
not search for the solution of  Eq. \eqn{eik2} that describes a spherical
wave originating from $P_S$, but we shall rather focus our attention on the
corresponding rays $\vec{x}(v)$. We observe that Eq. \eqn{eik2} has the
same form as the Hamilton-Jacobi equation for a unit mass particle, moving
in ${\bf R}^3$ with Hamiltonian:
\be
\bar{H}(x^i, l_i)= \frac{1}{2}(l_x^2+l_y^2+l_z^2)  + U(x,y,z)~,\label{hamil2}
\ee
where
\be
U(x,y,z)=2\Phi/c^2~.
\ee
Thus, the rays $\vec{x}(v)$ are solutions of the Hamilton's equations:
\be
\frac{d x^i}{dv}=\frac{\partial \bar{H}}{\partial l_i}~~~,~~
\frac{d l_i}{dv}=-\frac{\partial \bar{H}}{\partial x^i}~,~~~~~i=1,2,3\label{hameq}
\ee
subject to the constraint $\bar{H}=1/2$.

We imagine that, before reaching the bottom face $\Sigma_0$ of the
cylinder, the wave traverses a region of space where the gravitational
field is negligible and so the incoming wave entering the cylinder will not
differ appreciably from a spherical wave with center in $P_S$. If $Q$ is a
point of $\Sigma_0$ of coordinates $(\zeta,\xi,0)$, the eikonal $\bar
S(\zeta,\xi,0)$ will be roughly equal to the distance ${\rm
R}=[(\zeta-a)^2+
\xi^2 + l^2]^{1/2}$ of $Q$ from the source. We shall focus our attentions
on the rays that cross $\Sigma_0$ in directions approximately parallel to
the axis of the cylinder and this requires
\be
\frac{(\zeta-a)^2+\xi^2}{l^2} \ll 1.\label{para}
\ee
The weak-field condition, $\Phi/c^2(x,y,z)
\ll 1$, ensures that the direction of these rays
will continue to be practically parallel to the $z$-axis  all along the
cylinder and this allows us to use the paraxial approximation to study
their propagation. We start by taking the following Taylor expansion of the
eikonal of the incoming wave:
\be
\bar{S}|_{\Sigma_0}(\zeta, \xi)=  \; {\rm R} \approx \left(l +
\frac{(\zeta-a)^2+\xi^2}{2l}\right).
\label{incom}
\ee
From it we compute the components $l_x(\zeta,\xi)$ and $l_y(\zeta,\xi)$ of
the gradient of $\bar{S}(x)$ on the points of $\Sigma_0$. The third
component $l_z(\zeta,\xi)$ should be computed from the constraint ${\bar
H}=1/2$ (of course, we have to pick up the positive solution for $l_z$, as
our wave propagates in the positive $z$-direction) but due to the paraxial
condition $l_x(\zeta,\xi)\ll 1,\; l_y(\zeta,\xi) \ll 1$, and to the
weak-field condition $\Phi/c^2
\ll 1$, we can make the approximation $l_z(\zeta,\xi)=1$.
In conclusion, we can say with a good approximation that on $\Sigma_0$ the
rays $\vec{x}(v)$ associated with the spherical wave from $S$ are such,
that
$$
x(0;\; \zeta,\xi)=\zeta~,~~~~y(0;\; \zeta,\xi)=\xi~,~~~~~z(0;\;
\zeta,\xi)=0~, $$
\be
l_x(0;\; \zeta,\xi) =
\frac{\zeta-a}{l},~~~~l_y(0;\; \zeta,\xi)=\frac{\xi}{l} ,~~~~l_z(0;\; \zeta,\xi)= 1~
,\label{data}
\ee
where we have labelled each ray by the coordinates $(\zeta,\xi)$ of the
point, where it crosses $\Sigma_0$, and we have arbitrarily chosen $0$ for
the value of the parameter $v$ at that point.  Thus, in order to see how
the wave propagates inside the cylinder, in a small solid angle parallel to
the axis, we just have to find the solutions of the Hamilton's equations
\eqn{hameq}, using Eqs. \eqn{data} as initial conditions. Consider now the
case that the cylinder is much longer than thicker, $L
\gg D$ and that the distribution of dust inside it is independent of $z$ so
that, away from the caps, the potential $\Phi$ will be only a function of
$(x,y)$. The $z$-components of Eqs. \eqn{hameq} are then trivial to
integrate and give $z(v)=  v~$ and $l_z(v)=1$. We can thus use $z$ as a
parameter in the place of $v$. As for the projections of Eqs. \eqn{hameq}
in the $(x,y)$ plane, they are formally identical to the equations of
motion for a point particle on a plane, having a mass equal to one and
moving in the potential $U(x,y)$. It is obvious that if the potential
$U(x,y)$ has the shape of a two-dimensional well, there may exist a subset
$\Sigma^*
\subset
\Sigma_0$, such that for $(\zeta,\xi)\in
\Sigma^*$, the solutions $(x(z;\;
\zeta,\xi),~y(z;\;\zeta,\xi))$
represent bounded motions and, when this happens, it means that the rays
that cross $\Sigma^*$ will get trapped inside the cylinder. We thus have
that the gravitational field of the cylinder may effectively act as a sort
of waveguide for the gravitational radiation.  In order to examine this
phenomenon in greater detail, let us make the simplifying assumption that
the density $\rho_0$ of the dust is uniform. Then $U(x,y)$ inside the
cylinder is equal to the potential of a harmonic oscillator:
\be
\label{har}
U(x,y) = \frac{1}{2} \omega^2 (x^2 + y^2)~~~~~{\rm for}~~(x^2 + y^2)
\le D^2~,\label{o}
\ee
where
\be
\omega^2= \frac{4 \pi G \rho_0}{c^2}~.
\ee
If $D$ is infinite,  the motions $(x(z;\;\zeta,\xi),~y(z;\;\zeta,\xi))$ are
all bounded and thus all the rays will be trapped by the cylinder. Upon
solving the Hamilton's equations, we get
\beqa
x(z;\;\zeta,\xi)&=&\zeta\left[\cos (\omega z)+\frac{1}{\omega l}\sin(\omega
z)\right]~-\frac{a}{\omega l}\sin(\omega z),\nonumber\\
y(z;\;\zeta,\xi)&=&\xi\left[\cos (\omega z)+\frac{1}{\omega l}\sin(\omega
z)\right]~.\label{solray}
\eeqa
Let now $L_{\rm{foc}}$ be the solution of the transcendental equation:
\be
\tan(\omega L_{\rm{foc}})= -l \omega~~~~
\pi/2<\omega L_{\rm{foc}}<\pi~.\label{lfoc}
\ee
Assume that $L> L_{\rm{foc}}$ and let $N$ be the integer such that $N
L_{\rm{foc}}<L<(N+1)L_{\rm{foc}}$. We see from Eqs. \eqn{solray} that the
rays are focussed at the focal points $F_n$ of coordinates
$$
F_n\equiv \{-(-1)^n a \cos(\omega L_{\rm{foc}}),\; 0,\; L_{\rm{foc}}+
(n-1)\frac{\pi}{\omega}
\}~,
$$
\be
~~n=1,2,\cdots N
\ee
This means that inside the cylinder there will form a series of
alternatively inverted and right images of the source. We shall now compute
the specific luminosity ${\cal L}_n$ of the image at $F_n$, assuming that
the source $S$ has a specific luminosity ${\cal L}_S$. According to the
definition of ${\cal L}_S$, and assuming that $S$ radiates isotropically,
the number $dN_S$ of gravitons with energy in the range $\hbar d
\omega_s$ emitted by $S$ during the proper time interval $d
\tau_S$ in the solid angle $d
\Omega_S$ around the $z$-axis is equal to
\be
dN_S=d\tau_S \,d \Omega_S \,d\omega_S\, \frac{{\cal L}_S}{4 \pi \omega_S}~.
\ee
These gravitons form a narrow bundle. According to what we said above, they
will be focussed at $F_n$, and we let $d \Omega_n$ be the solid angle they
span at $F_n$. By the graviton number conservation, Eq. \eqn{cons}, $d N_S$
must be equal to the number of gravitons $dN_n$ emitted by the image during
the corresponding time interval $d\tau_n$, in the interval of frequency $d
\omega_n$ in the solid angle $d \Omega_n$:
\be
dN_S=dN_n=d\tau_n \,d \Omega_n \,d\omega_n\, \frac{{\cal L}_n}{4 \pi
\omega_n}~.
\ee
Since the source and the image are at rest, $d \tau_S=d \tau_n$,
$\omega_S=\omega_n$, and $d \omega_S=d\omega_n$. Thus, we have
\be
{\cal L}_n ={\cal L}_S \left\vert \frac{d \Omega_S}{d
\Omega_n}\right\vert~.
\ee
In view Eqs. \eqn{solray}, it is easy to verify that
\be
\frac{d \Omega_n}{d \Omega_S}= \cos^2(  \omega L_{\rm{foc}})(1+\omega^2 l^2)^2~.
\label{lum}
\ee
If $\omega l \ll 1$ we see from Eq. \eqn{lfoc} that $L_{\rm{foc}}\approx
\pi/\omega
- l$. Equation \eqn{lum} then implies ${\cal L}_n \simeq {\cal L}_S(1-\omega^2 l^2)
\approx {\cal L}_S$, which shows that the images have practically the same specific luminosity of
the source. From the point of view of the observer $O$, it will be as if
the ``waveguide" had driven the source from $P_S$ to $F_N$. Being much
closer to him than the real source, the image at $F_N$ will obviously
appear to him much brighter.

So far, we have proceeded as if the radius of the cylinder had been
infinite. The effect of $D$ being finite is that in general not all the
rays entering the cylinder will be focussed at $F_N$. Outside the cylinder,
the strength of the gravitational field decreases rapidly as one goes
farther from its axis and so in general a ray that escapes out of it will
not be focussed at $F_N$. This suggests that a simple criterion for a ray
to be focussed is that it should always remain inside the cylinder while
traversing it: $x(z)^2+y(z)^2< D^2$ for $0< z< L$.  Assuming again that
$\omega l \ll 1$, we see from Eqs. \eqn{solray} that this condition is
satisfied only by the gravitons emitted by $S$ in a solid angle $\Delta
\Omega_S$ around the direction of the $z$-axis of order:
\be
\Delta \Omega_S \approx \frac{ D }{L_{\rm foc}}\pi~.
\ee
Since, for $\omega l \ll 1$, this is also roughly equal to the angle
$\Delta \Omega_N$ spanned by these rays at the focal point $F_N$, $\Delta
\Omega_S$ provides a measure of the degree of alignment of $O$ with the axis
of the cylinder needed for $O$ to observe the image at $F_N$.

In order to have an estimate of the above numbers in realistic situations,
let us imagine that our waveguide is constituted by a sequence of galaxies
forming a filamentary structure of the kind observed in the large scale
structure of the Universe. Taking for $D$ the typical size of a galaxy, $D
\approx 10^4$ pc, and for $\rho_0$ the typical density of matter inside a
galaxy, $\rho_0\approx 5\cdot 10^{-24}$ g/cm$^{3}$, we find that the
weak-field approximation for the background, the second inequality in Eq.
\eqn{weak}, is satisfied, because near the edge of the cylinder, where the
galactic gravitational field is strongest, $2\Phi/c^2\ \approx
\omega^2 D^2/2 \approx 10^{-6}$. Plugging the above value for $\rho_0$ in Eqs. \eqn{o}
and \eqn{lfoc}, we get that $L_{\rm foc} \approx 10^2$ Mpc, which is the
typical scale of the large structures of the Universe and represents also
the expected distance of coalescence events, as was discussed in the
Introduction. The corresponding angular opening $\Delta
\Omega_S$ of the bundle of rays that are efficiently trapped is of the
order of $1'$.

In the above discussion, we have treated our waveguide as if the
distribution of matter inside it was homogeneous. Obviously this is not
what happens inside the galaxies, where a significant fraction of matter is
concentrated in compact bodies like the stars. \footnote{Considering the
recent results of microlensing collaborations like MACHO, OGLE, DUO, and
EROS, it is very likely that a large amount of the matter inside galaxies
is built up of baryonic compact objects weakly interacting with
electromagnetic radiation like planets, brown dwarfs or black holes so that
all the following considerations apply also to these galactic components
\cite{paczynski},\cite{bartelmann},\cite{schneider},\cite{ehlers}.} The
large scale galactic field is distorted on small scales by the local field
of nearby stars and we have to analyze the effect of these local
disturbances on the focussing properties of the waveguide. A simple
order-of-magnitude estimate shows that in fact it should be negligible. We
start by computing the distance $d(x)$ such that a ray, while traversing
the cylinder, passes with probability $x$ at a distance smaller than $d(x)$
from a star. For this purpose, consider the tube ${\cal T}$ of radius
$d(x)$ around the ray. Assuming that the ray is roughly parallel to the
axis of the cylinder and that the stars are distributed in the cylinder
with uniform density $n_*$, the probability of finding a star inside ${\cal
T}$ is
\be
x=1-(1-d^2(x)/D^2)^{n_* \pi D^2 L}\simeq n_* \pi d^2(x) L~.
\ee
This formula is easy to understand, since $(1-d^2(x)/D^2)^{n_* \pi D^2 L}$
is the probability that throwing at random $n_* \pi D^2  L$ stars inside
the cylinder no one of them falls inside ${\cal T}$.  Now, since the
deflection $\alpha(x)$ suffered by a ray passing at a distance $d(x)$ from
a star with mass $M$ is approximately $\alpha(x)\simeq 4 G M/c^2 d(x)$, we
have
\be
\alpha(x)\simeq \frac{4 G}{c^2} \sqrt{\frac{\pi M^2 n_* L}{x}}~.
\ee
Taking $L\approx 10^8$ Mpc, $M n_*\simeq 0.12
\;M_{\odot}$pc$^{-3}$ \cite{all} and assuming that the stars have a mass of
the order of the mass of the sun $M_{\odot}$, we get
\be
\alpha(x)\simeq 2 \times 10^{-4} \frac{1}{\sqrt x}\, {\rm arcsec}.
\ee
For $x =0.01$, the deflection angle is of order of $10^{-3}$ arcsec. This
means that, while traversing the cylinder from one end to the other, a ray
has a probability of $1\%$ of being deflected by a star by an angle of
approximately one thousandth of arcsecond. Now consider the bundle of rays
emitted by $S$ in a solid angle of opening $\Delta
\Omega_S$. While traversing the cylinder, the cross section of this bundle
will expand up to a point where it becomes comparable to the cross section
of the cylinder and then, after traversing a distance of order $L_{\rm
foc}$, it will shrink to a point. The rays in the bundle will thus suffer
an average deflection of order $D/L_{\rm foc}\simeq 1'$, which is much
greater than the deflection due to the occasional passages near a star as
estimated above. We thus conclude that the stars have a negligible effect
on the propagation of the rays inside the cylinder.

\section{Discussion and conclusions}

In this paper, we have shown that a filamentary distribution of dust may
act as a waveguide for gravitational radiation. We derived this result
studying the short wave-length limit of the linearized Einstein's equations
that describe the propagation of a weak gravitational wave through dust. In
this approximation, a gravitational wave behaves in essentially the same
way as an electromagnetic wave propagating in vacuum: in particular, the
gravitational wave propagates along geodesics of the background
gravitational field generated by the dust. The dust itself behaves as a
non-dispersive and non-dissipative medium, preserving the polarization
state of the gravitational wave. In this geometrical optic limit, the
waveguiding effect has a very simple explanation: the gravitational field
generated by a filament of dust traps in its interior a paraxial bundle of
rays, with the result that, while traversing the filament, the amplitude
$h$ of the gravitational wave does not undergo the $1/R$ attenuation
characterizing the propagation in vacuum of spherical waves. We showed also
that for a roughly uniform density of dust, the guide produces real images
of the source, having approximately the same absolute ``luminosity" as the
real source. In some sense, the waveguide "draws" the source closer to the
observer: if the true distance to the source is $R$, its image brightness
will correspond to that of a similar source at the closer distance
\be
R_{eff}=R-l-L\,,
\ee
where $L$ is the length of the filament and $l$ is the distance of the
source from the top of the filament (see also \cite{cap}).

Even though the effect described in this paper is basically a gravitational
lensing effect, it is worth pointing out that it differs in some important
points from the standard gravitational lensing situations. When discussing
the gravitational lensing of light, it is normally sufficient to treat the
deflector as a thin lens; moreover it is usually  assumed that the
gravitational field of the deflector can be treated using the newtonian
limit, which requires the impact parameter $r_{0}$ to be much larger than
the gravitational $r_{g}$ radius of the lens $r_{0}\gg r_{g}$. This is not
a very stringent constraint, because of the smallness of $r_{g}$; a much
more stringent condition is that the light ray should pass close enough to
the lens, for the deflection to be sensible, but not so close to cross it,
because this would easily lead to the absorption or scattering of the light
ray by the matter constituting the lens. In order to obtain the waveguiding
effect described in this paper, we need to reverse two of the three above
conditions: while we still assume the weak field condition, which is well
respected in realistic astrophysical situations, an efficient trapping of
the gravitational wave requires the lens to be very thick (in fact our
filamentary lenses are much thicker than wider), and that the wave passes
well inside the distribution of matter, in the region where the newtonian
potential well is deeper and thus more confining. These conditions do not
represent a problem since gravitational waves, differently from
electromagnetic waves \cite{pirani},\cite{straumann}, travel nearly
unscathed through all forms and amounts of intervening matter \cite{thorne}
and can thus traverse our long filaments without being significantly
absorbed or scattered. Finally, we should mention that even though in this
paper we have limited the discussion of the waveguiding effect to a dust
distribution with the shape of a filament, more general geometries, planar
for example, analogous to those used for optical waveguides, could be
considered.

If the waveguiding effect described in this paper could be really observed,
it might significantly lower the sensitivity threshold required to
successfully detect a gravitational wave. In order to address this issue
several questions have to be answered. The first one is: what is the
probability that waveguides for gravitational radiation exist? Furthermore,
what is the probability that a significant fraction of the flux from
sources of gravitational radiation will get trapped?

The first point seems, from our point of view, very likely. The
observations in the last twenty years tell us that filamentary or slab-like
structures are quite common at large scales (see, e.g.
\cite{peebles},\cite{fort},\cite{kolb},\cite{huchra}). Furthermore, their
sizes ($\sim 10^2\div 10^3$ Mpc) are comparable with the focalization
length of waveguides, as we have shown above. The hypotheses of their
origin, starting from some fundamental theory and in connection with
primordial structures as cosmic strings, domain walls and textures are
among the most investigated topics of theoretical cosmology
\cite{kolb},\cite{turner},\cite{vilenkin},\cite{gott}.

The second point depends on the efficiency of production of gravitational
radiation. As we said in the Introduction, for rates of the order of a few
events per year, the expected distance of high frequency sources like
coalescing neutron-stars, should be in the range of $10^2\div 10^3$ Mpc,
which is also the order of magnitude of the focalization length of our
waveguides. If the probability that a wave from such a source gets trapped
by a waveguide is large enough, this effect might be seen already by the
VIRGO and LIGO interferometers in the next few years. Since the filamentary
structures that constitute our waveguides extend over cosmological
distances, the waveguiding effect may have an important impact on the
detection of gravitational waves in the low-frequency band that will be the
target of the LISA interferometer and the candidates for such a kind of
sources are quite numerous. First of all we have to stress the fact that,
obviously, all the sources we are considering have extragalactic origin and
can  be sited in distant galaxies. This fact implies that our sources may
have high redshifts so that in the frequency and in all the quantities
characterizing a gravitational wave we have to include a factor $(1+z)$. A
large amount of candidates sources could be formed by binary evolved stars
as double systems of white dwarfs, black holes and white dwarfs, or black
holes and black holes. Such structures are very common in the Universe,
they are present in all galaxies and, if they are tight bound, the
gravitational radiation emission fluxes are conspicuous. Another class of
objects could be quasars \cite{arp} if in their core  a very massive black
hole is present and massive objects fall into it. Finally, waves from
primordial phase transitions and from cosmic strings \cite{vilenkin} could
be good candidate sources.



\vfill

\end{document}